\documentclass[aps,twocolumn,
showpacs,superscriptaddress,citeautoscript,prb,floatfix, 10pt]{revtex4-2}
\usepackage{bm}
\usepackage{graphicx}
\usepackage{amsmath,amssymb}
\usepackage{amsfonts}
\usepackage{color}
\usepackage{siunitx}
\usepackage{physics}
\usepackage{bbm}
\usepackage{hyperref}

\definecolor{Red}{rgb}{0.9,0.3,0.3}

\begin{document}

\title{Unconventional Floquet topological phases in the SSH lattice}

\author{Dunkan Mart\'inez }
\affiliation{GISC, Departamento de F\'isica de Materiales, Universidad Complutense de Madrid, Spain}	
\author{Yuriko Baba}
\affiliation{Condensed Matter Physics Center (IFIMAC), Universidad Aut\'onoma de Madrid, E-28049 Madrid, Spain}
\affiliation{GISC, Departamento de F\'isica de Materiales, Universidad Complutense de Madrid, Spain}
\author{Benjam\'in Santos}
\affiliation{M82 Consulting Services G.P., Laval, Canada}
\author{Rodrigo P. A. Lima}
\affiliation{GISC \& InterTEP, Departamento de F\'{\i}sica, 
Facultad de Ciencias Ambientales y Bioqu\'{\i}mica, Universidad de
Castilla-La Mancha, 45071 Toledo, Spain}
\affiliation{GFTC, Instituto de F\'{\i}sica, Universidade Federal de Alagoas, Macei\'{o} AL 57072-970, Brazil}
\author{Pedro Orellana}
\affiliation{Universidad Federico Santa Mar\'ia, Chile}
\author{Francisco Dom\'inguez-Adame}
\affiliation{GISC, Departamento de F\'isica de Materiales, Universidad Complutense de Madrid, Spain}	
\author{Alexander L\'opez}
\email[To whom correspondence should be addressed. Electronic
address: ]{alexlo08@ucm.es}
\affiliation{GISC, Departamento de F\'isica de Materiales, Universidad Complutense de Madrid, Spain}
\affiliation{Institute for Theoretical Physics, University of Regensburg, D-93040
Regensburg, Germany }
\affiliation{Escuela Superior Polit\'ecnica del Litoral, ESPOL, Departamento de F\'isica, Facultad de Ciencias Naturales y Matem\'aticas, Campus Gustavo Galindo
 Km. 30.5 Via Perimetral, P. O. Box 09-01-5863, Guayaquil, Ecuador}


\begin{abstract}

Topological materials, known for their edge states robust against local perturbations, hold promise for next-generation quantum technologies, but remain scarce in nature and challenging to realize in static systems. The Su–Schrieffer–Heeger chain  is a one-dimensional system for topological phases, although its static control is limited.  To overcome these limitations, we propose to use high-frequency monochromatic driving and modulated amplitude pulses to dynamically induce and switch the Floquet topological phases. Using a Kramers–Henneberger–like transformation, we encode all Floquet sidebands into a single effective Hamiltonian. We demonstrate that both monochromatic and experimental pulse protocols (Gaussian and fast-beating envelopes) can induce topological edge states, enabling dynamic phase switching. Notably, fast-beating modulations require significantly lower field strengths than monochromatic ones, especially with larger inter-dimer separations. Our findings offer an experimentally feasible route for Floquet engineering, paving the way for ultrafast and energy efficient control of topological phases in quantum platforms, opening up new possibilities in the field of dynamic quantum materials.


\end{abstract}
\maketitle

----------------------------------------------------------------%
    
    

\section{Introduction}\label{section1}

Topological materials have received a great deal of attention since their discovery~\cite{VonKlitzing1980,Haldane1988,Kane2005,Bernevig2006,Koenig2007,Hasan2010}. Their engineering via periodically driving time-dependent interactions has been shown to be a suitable means to manipulate material properties on demand. These topological materials can support a bulk bandgap, but have protected gapless edge states at their boundaries.  Spinless systems have been shown to possess protected edge states in the absence of time-reversal symmetry~\cite{Bernevig2014}.
In pioneering works~\cite{Oka09,Dora2009,Lindner2011,usaj2014,Grushin2014,Titum2017,Esin2018,Peng2019,Lu2019}, it is shown that non-trivial topological properties can be induced in static systems by periodically driving fields, leading to the so-called Floquet topological insulators (FTI), which show boundary edge states that are not achievable within the static scenario. Generally speaking, the topological response of a physical system can be assessed via topological invariants, such as the Chern number, the spin-Chern number $\mathcal{Z}_2$, and higher-order extensions. The ten-fold classification and its extensions allow one to determine which class belongs to a given system solely by its symmetries and spatial dimension~\cite{Kitaev2009,Schnyder2008,Chiu2016,Chiu2013,Langbehn2017}. An interesting model for realizing FTIs phases in periodically driven lattices, valid for arbitrary driving regimes, is presented in ~\cite{Platero2013}, demonstrating that nonadiabaticity drives transitions between distinct topological phases. 

Remarkably, several works on FTIs focus on constant-amplitude periodic interactions. However, maintaining a constant-amplitude electric field can be challenging in experimental setups, whereas the use of finite-size pulses modulated by amplitude and frequency, achievable within current technological implementations, allows for a more realistic description of the actual experimental scenario~\cite{Baba_2025}. For example, recent advances in time- and angle-resolved photoemission spectroscopy have enabled access to ultrafast electron states and their spin dynamics in solids~\cite{Ito2023, McIver2019, Merboldt2025, Choi2025}. 
From the theoretical point of view, the description of such experimental scenarios needs to take into account the non-periodicity of the systems, i.e. regimes beyond conventional Floquet engineering. This can be done with direct solutions to the time-dependent problem or with a hybrid formalism such as the so-called $t-t'$ formalism  that describes the effects of the driving field using two distinct time scales, namely the envelope amplitude time scale and the time period of the external field, which is described within a Floquet formalism~\cite{Drese1999, Holthaus2015, Ikeda2022, Baba_2025, Alejandro_2025}. Even if the $t-t'$ formalism produces a rather simple extended Floquet picture and brings a powerful tool for interpreting dynamics under pulsed ultrafast periodic driving, when dealing with short pulses the formalism fails as it considers that in the time period of the external field the envelope amplitude remains almost constant. Furthermore, when dealing with intense radiation fields, the need for several sidebands to achieve convergence within the Fourier expansion is computationally intensive. 
This could be particularly challenging in finite-size systems, where it is desirable to have a less computationally demanding method that correctly describes the dynamical transitions among the static energy states induced by the driving, and their related observables. 
In particular, in the system we address, it is crucial to quantify observables such as the population measuring the average probability of occupation of the generated topological states. 

In this work, we show that intense pulses generate unconventional non-trivial Floquet topological states whose populations can be tuned on demand. To this end, we implement an alternative strategy to deal with arbitrary pulse shapes by means of a Krammers-Henneberger-like transformation~\cite{Henneberger}, which we apply to study a periodically driven one-dimensional Su-Schrieffer-Heeger (SSH) chain~\cite{Su1979,Li2017,dalLago2015floquet,Borja2022}. Interestingly, although the topological phases in the static SSH model are broken by the inclusion of static chiral symmetry-breaking terms in the electron Hamiltonian, we show that for a dipolar periodic driving protocol, despite instantaneously breaking the chiral symmetry, robust Floquet edge states are still induced even for parameters where the static system remains trivial. Moreover, the amplitude-modulated protocols enable the realization of these non-trivial topological phases at lower effective values of the light-matter coupling strength, compared to the values required for constant-amplitude modulations. We expect these features to be advantageous for actual experimental realizations of our theoretical proposal. The structure of this work is as follows. In Sec.~\ref {section2} we apply the Krammers-Henneberger transformation to effectively solve the dynamical response of a one-dimensional SSH chain subject to a periodically driving electric field with constant amplitude, showing the population tunability of the quasi-energy spectrum. Then, in Sec.~\ref{sec:driving} we extend the analysis to amplitude-modulated electric field configurations, comparing explicitly Gaussian and harmonic-modulated pulses, which constitute two experimentally relevant driving protocols. Section~\ref{sec:conclusions} presents the concluding remarks.

\section{Periodically driven SSH chain}\label{section2}

\begin{figure}
    \centering
    \includegraphics[width=0.9\linewidth]{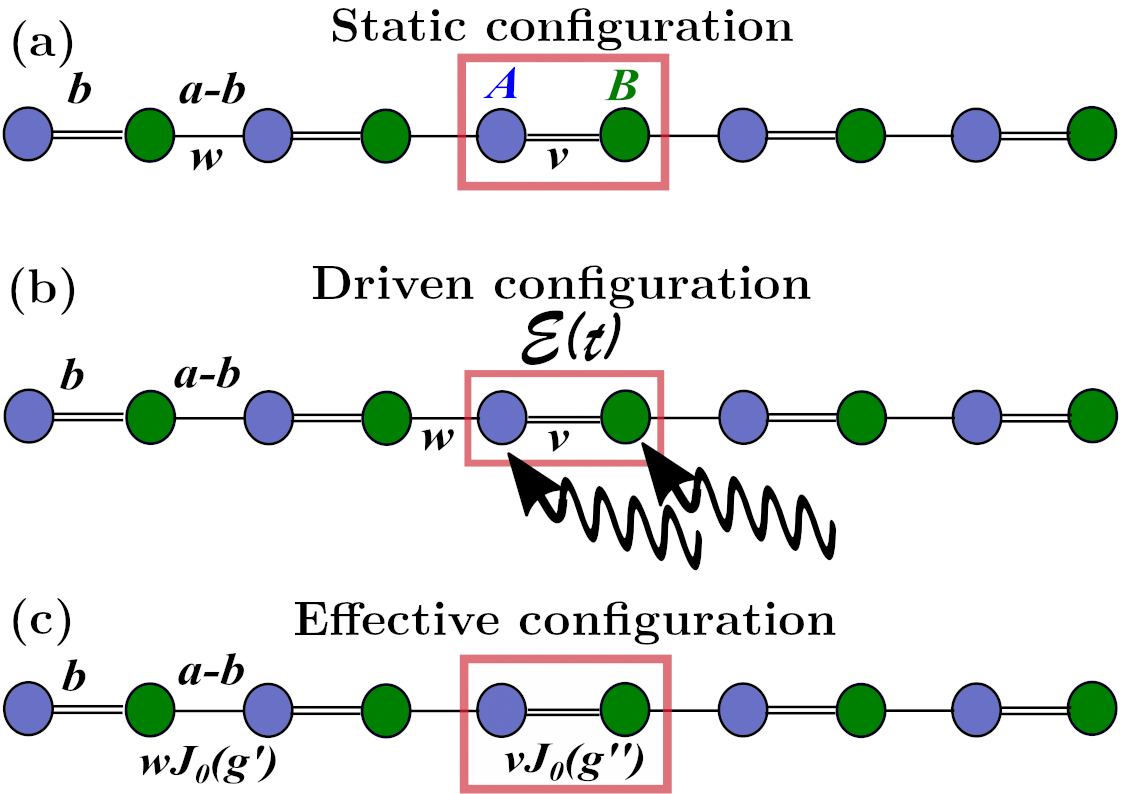}
    \caption{Schematic representation of the (a)~static, (b)~driven and (c)~effective one-dimensional SSH chain. The lattice and intracell dimer separations are denoted as $a$ and $b$, respectively. The unit cell is highlighted within the red rectangle. The intracell and intercell hopping energies are $v$ and $w$, repectively, whereas $\mathcal{E}(t)$ is the time-dependent electric field along the chain and the effective light-matter coupling strength are $g^{\prime}=(1-b/a)g$ and $g^{\prime\prime}=gb/a$ with $g=ea\mathcal{E}_0/\hbar\Omega$.}
    \label{fig:Sketch}
\end{figure}

We consider spinless electrons in a periodically driven one-dimensional SSH chain with $N$ dimers, with staggered hopping energies, and subject to a time-dependent uniform electric field, as shown schematically in Fig.~\ref{fig:Sketch}. The static Hamiltonian for the bipartite system is given as~\cite{Platero2013,Li2017}
\begin{multline}
H_S= v\sum_{\ell=1}^N \ket{\ell}\bra{\ell}\sigma_x + w \sum_{\ell=1}^{N-1}\ket{\ell+1}\bra{\ell}\sigma_{+} + \mathrm{h.c.}\ ,
\end{multline}
where $\ket{\ell}$ represents the two orbitals located in the $\ell\,$th unit cell and $\sigma_\pm=(\sigma_x\pm i\sigma_y)/2$ are the usual ladder operators, with Pauli matrices acting in the inner space of the $\{A,B\}$ sublattices. In addition, $v$ and $w$ denote the intra- and inter-cell hoping energies. This SSH Hamiltonian was proposed to theoretically describe solitary wave states in the trans poly-acetylene molecule \cite{Su1979}. 

When the chain is subject to a periodic driving protocol, via a uniform monochromatic electric field directed along the chain and described within the dipolar approximation, the total time-dependent Hamiltonian $H(t)=H_S+V(t)$ reads
\begin{equation} \label{eq:Ht}
H(t)=H_S+e\mathcal{E}(t)x\ ,    
\end{equation}
with $-e$ the electric charge of the electron, $\mathcal{E}(t)$ the time-dependent electric field, and $x$ the position operator with respect to the center of the chain. Within this formulation, the driving interaction is symmetric with respect to the reflection operation $x\rightarrow -x$.  In the tight-binding approach, the interaction takes the form
\begin{equation}
V(t)=e\mathcal{E}(t)\sum_{\ell=1}^N\left[\frac{(2\ell - N - 1)}{2}a\mathbbm{1}_\sigma  - \frac{b}{2}\,\sigma_z\right]\ket{\ell}\bra{\ell} \ ,
\label{eq:potential_basis}
\end{equation}
with $a$ being the lattice spacing and $b$ the distance between the sublattice atoms~\cite{Platero2013}. Here, $\mathbbm{1}_\sigma$ is the identity operator in the degree of freedom of the sublattice. The last term in Eq.~\eqref{eq:potential_basis} shows that whenever the inter-dimer separation $b$ is nonzero,  the electric field dynamically breaks the chiral symmetry of the system given by the operator $\sigma_z$. It is well-known that in the static regime, without chiral symmetry, the SSH chain cannot sustain topological phases. 
The Schr\"odinger equation for the time-dependent Hamiltonian~\eqref{eq:Ht} is 
%
\begin{equation}
i\hbar\partial_t\ket{\Psi(t)}=H(t)\ket{\Psi(t)}~,
\end{equation}
and its solution can be written as a linear combination of the eigenstates of $H_S$ 
\begin{equation}
\ket{\Psi(t)}=\sum_{\ell=1}^N\sum_{s=\pm}c_{\ell,s}(t)\ket{\ell}\ket{s}\ .
\end{equation}
where $\ket{s}$ is the eigenvector related to the sublattice space.

To solve the dynamical evolution equation, we first perform a unitary transformation $\ket{\Psi(t)}=P(t)\ket{\Phi(t)}$ into the time-dependent Schr\"odinger equation, where $P(t)P^\dagger(t)=\mathbbm{1}_\sigma\mathbbm{1}_N$, with $\mathbbm{1}_N$ being the N-dimensional identity operator. Therefore, we get the transformed Hamiltonian
\begin{equation}
\mathcal{H}(t)=P^\dagger(t)H(t)P(t)-i\hbar P^\dagger(t)\partial_tP(t)\ .    
\end{equation}

Up to now, the result has been general. Now we choose
\begin{equation}
i\hbar \partial_tP(t)=V(t)P(t)\ , 
\end{equation}
whose solution is given by
\begin{equation}
P(t)=e^{iB(t)\sigma_z/2}\sum_{\ell=1}^Np_\ell(t)\ket{\ell}\bra{\ell}\ ,    
\label{eq:transf_gen}
\end{equation}
where $p_\ell(t)=\exp[-i(a/b)(\ell-N/2-1/2) B(t)]$ and $B(t)=(eb/\hbar)\int^t  \mathcal{E}(t^\prime)dt^\prime$. Since this unitary transformation also satisfies $[P(t),V(t)]=0$, the transformed Hamiltonian then reads
\begin{equation} \label{eq:H_transf}
\mathcal{H}(t)=P^\dagger(t) H_S P(t)\ .    
\end{equation}

The effective time-dependent Hamiltonian is given by:
%
\begin{align}
\mathcal{H}(t) &= v \sum_{\ell=1}^N \ket{\ell}\bra{\ell} \Big(e^{-iB(t)} \sigma_+ + \mathrm{h.c.}\Big) \nonumber \\
&+ w \sum_{\ell=1}^{N-1} \Big[p(t) \ket{\ell+1}\bra{\ell} e^{-iB(t)}\sigma_+ + \mathrm{h.c.}\Big]\ .    
\label{eq:General}
\end{align}
where $p(t)=\exp[i(a/b)B(t)]$.
For the effective electric field $\mathcal{E}(t)=\mathcal{E}_0\cos\Omega t$, we get
\begin{align}
\mathcal{H}(t) &= v \sum_{\ell=1}^N \ket{\ell}\bra{\ell} \Big(e^{-irg\sin\Omega t}\sigma_+ + \mathrm{h.c.}\Big) \nonumber \\
&+ w\sum_{\ell=1}^{N-1} \Big[\ket{\ell+1}\bra{\ell} e^{i(1-r)g\sin\Omega t} \sigma_+ + \mathrm{h.c.}\Big]\ ,   
\label{eq: Heffective}
\end{align}
%
where $g=ea\mathcal{E}_0/\hbar\Omega$ is the effective light-matter coupling and $r=b/a$ is the distance ratio.

The Floquet theorem states that the solution of a time-periodic Hamiltonian can be expressed as a linear combination of functions sharing the same periodicity~\cite{Shirley1965,Grifoni1998}
\begin{equation}
   \ket{\Psi(t)} = \sum_\alpha f_\alpha \phi^F_\alpha(t) \, .
\end{equation}
Here $\phi_\alpha^F(t) = e^{-iE_\alpha t/\hbar} u_\alpha(t)$ is called the Floquet function and can be expressed as a periodic function $u_\alpha(t)$ modulated by a non-periodic part depending on an exponent $E_\alpha$.  Since $u_\alpha(t)$ is periodic, we can expand it in its Fourier components 
\begin{equation}
    u_\alpha(t) = \sum_{m=-\infty}^\infty e^{im\Omega t} u_{\alpha,m}\, .
\end{equation}
Substituting this into the definition of the Floquet function gives
\begin{equation}
    \phi_\alpha^F(t) = e^{-i\xi_{\alpha, m} t/\hbar} u_{\alpha, m} 
 \end{equation}
 where $\xi_{\alpha, m} = E_\alpha -m\hbar \Omega$ is called the quasi-energy, stemming from the fact that, for a time-dependent Hamiltonian,  energy is not a conserved quantity (for further discussion, see~\cite{Giovannini_2020}. This decomposition can be used to find an equivalent formulation to the explicitly time-dependent Hamiltonian $\mathcal{H}(t)$ in terms of its Floquet-Fourier coefficients
%
\begin{equation}
H_{mm'}=\frac{1}{T}\int_0^T\mathcal{H}(t)e^{i(m-m')\Omega t}dt -m\hbar\Omega \delta_{mm'}\mathbbm{1}_{2N}\ , 
\label{Floquet-Fourier}
\end{equation}
where $\mathbbm{1}_{2N}=\mathbbm{1}_\sigma\sum_{\ell=1}^N\ket{\ell}\bra{\ell}$ is the  $2N\times2N$ identity matrix describing the dimer and lattice sites subspace for a chain with $N$ dimers. The pulse period is $T=2\pi/\Omega$. In practice, the Fourier series to build the Floquet-Fourier matrix is truncated for a maximum number of replicas $M_{max}$ so that the spectrum, state populations, and other physical quantities of interest do not change qualitatively upon introducing additional Floquet replicas.  Thus, the full Floquet-Fourier Hamiltonian can be written as
\begin{equation}
    \mathcal{H}_{FF} = \sum_{m,m'}^{M_{max}} H_{mm'} \ket{m} \bra{m'}
\end{equation} 
where $\ket{m}$ is the space of the replicas.

Since the effective electric field is a monochromatic radiation field with constant amplitude, we can make use of the Jacobi-Anger expansion $\exp( i z\sin\Omega t)=\sum_{n=-\infty}^{\infty} J_{n}(z)\exp(in\Omega t)$ in terms of the Bessel functions of the first kind,
%
%
%
to write the effective Hamiltonian~\eqref{eq: Heffective} in the space of the Floquet replicas. Using the orthogonality relation $(1/T)\int_0^Te^{i(m-m')\Omega t}dt=\delta_{mm'}$ we can write the Floquet-Fourier Hamiltonian \eqref{Floquet-Fourier} as
%
\begin{align}\label{Floquet-Fourier4}
& H_{mm'} = \sum_{\ell=1}^N \ket{\ell}\bra{\ell} \Big[v J_{n}(g^{\prime}) \Big((-1)^{n}\sigma_+ + \sigma_-\Big)\nonumber \\
& - m\hbar\Omega\mathbbm{1}_\sigma \delta_{mm'}\Big] + w J_{n}(g^{\prime\prime}) \nonumber \\
& \times \sum_{\ell=1}^{N-1}\Big[ (-1)^{n}\ket{\ell+1}\bra{\ell}\sigma_+ + \ket{\ell}\bra{\ell+1}\sigma_-\Big]\ . 
\end{align}
%
where $n = m'-m$, $g^{\prime}=rg$, $g^{\prime\prime}=(1-r)g$ and we have used the parity property of the Bessel functions $J_m(-z)=(-1)^mJ_m(z)$.
For large frequencies, i.e, when $w,v\ll\hbar\Omega$, one can approximate the Floquet-Fourier Hamiltonian as  $\mathcal{H}_{FF}\approx H_{00}$ with
%
\begin{align}
H_{00}&=vJ_0(g^{\prime})\sum_{\ell=1}^N\ket{\ell}\bra{\ell}\sigma_x+wJ_0(g^{\prime\prime})\nonumber \\
&\times \sum_{\ell=1}^{N-1}(\ket{\ell+1}\bra{\ell}\sigma_++\ket{\ell}\bra{\ell+1}\sigma_-)\ . 
\label{eq: HFloquet}
\end{align}
%
From Eq.~\eqref{eq: HFloquet} we can see that the inter-dimer separation $b$ effectively adds a modulation of the intercell hopping parameter $v$. Since Bessel functions are oscillatory bounded by $\max\big(J_m(x)\big)\le1$, when $b = 0$, only the intracell hopping $w$ is renormalized. Here, the topological state disappears when $wJ_0(g) < v$ while the trivial state remains unchanged. In contrast, when $b \ne 0$  the modulation extends to the intercell hopping as well, allowing for regions in $g$-space where $vJ_0(g^{\prime}) < wJ_0(g^{\prime\prime})$, thus enabling the emergence of a topological phase.

\begin{figure}[ht] 
\centering
\includegraphics[width=0.9\linewidth]{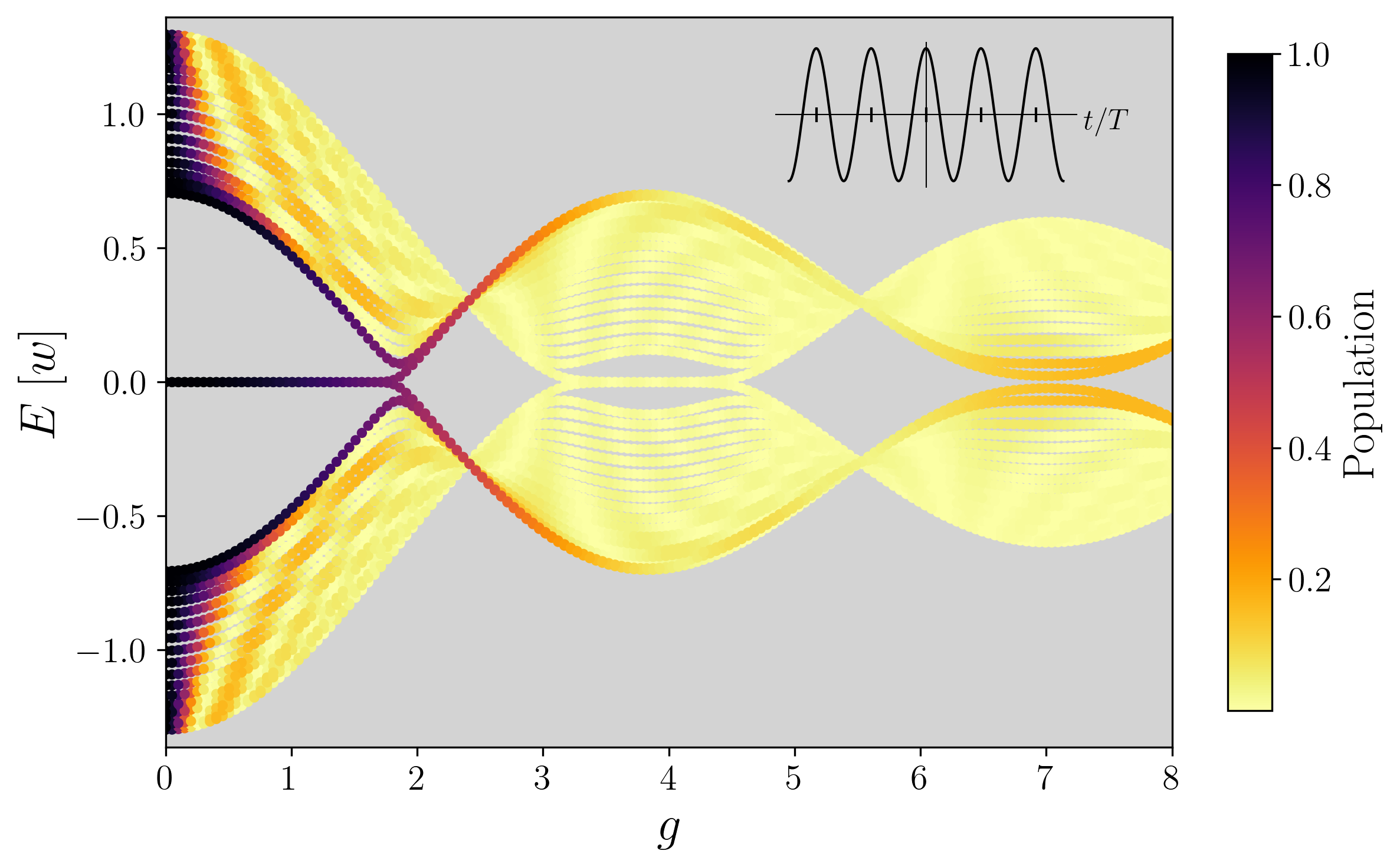}
\caption{Quasienergy spectrum within the off resonant regime $\hbar\Omega/w=10$, for the static topological phase $v/w=0.3$ and $b=0$. The color scale represents the state population associated with each quasi-energy state. The inset shows the time-dependent profile of the applied electric field, and the ticks in it's $x$-axis represent the integrating regions of Eq.~\eqref{Floquet-Fourier}.}
\label{fig:Trivial}
\end{figure}

For the numerical discussion, we have considered $w$ as our unit of energy. Therefore the topological state is characterized by $0\le v<1$. For a better comparison with the results reported in Ref.~\cite{Platero2013}, we have obtained the spectra and eigenstates of a $N=20$ SSH chain when it is driven by a high-frequency electric field (i.e. $\hbar \Omega = 10w$) in terms of the effective light-matter interaction strength $g$. Here, the squared modulus of the eigenstates can be interpreted as the time-dependent occupancy of the different Floquet replicas
\begin{equation}
        p_{\alpha, m} = \frac{1}{T}\int_0^T || P(t) u_{\alpha. m} ||^2 \,dt\ .
\end{equation}
In this way, we can spot when the gap opens or closes, and also quantify whether or not the topological states are populated. Using the unitary transformation, the quasienergy spectra can be obtained directly from~ Eq.~\eqref{eq: HFloquet} and it does not require the use of replicas for convergence. However, obtaining the population requires the use of at least $M = 20$ replicas. Figure~\ref{fig:Trivial} confirms the predicted behavior in which the trivial and topological states change with increasing $g$. However, we find that the next topological state is not populated, limiting its applicability in potential devices that rely on the topological properties of the system. Furthermore, the population exhibits an oscillatory behavior as $g$ increases, with an oscillation frequency that increases with energy. Although a direct numerical explanation for this effect is not evident, the structure of the unitary transformation used suggests that these oscillations can be traced back to the modulation of the wavefunction expressed through the Bessel functions, which naturally arise from the expansion of the time-periodic driving field.

\begin{figure*}[ht] 
\centering
\includegraphics[width=0.8\linewidth]{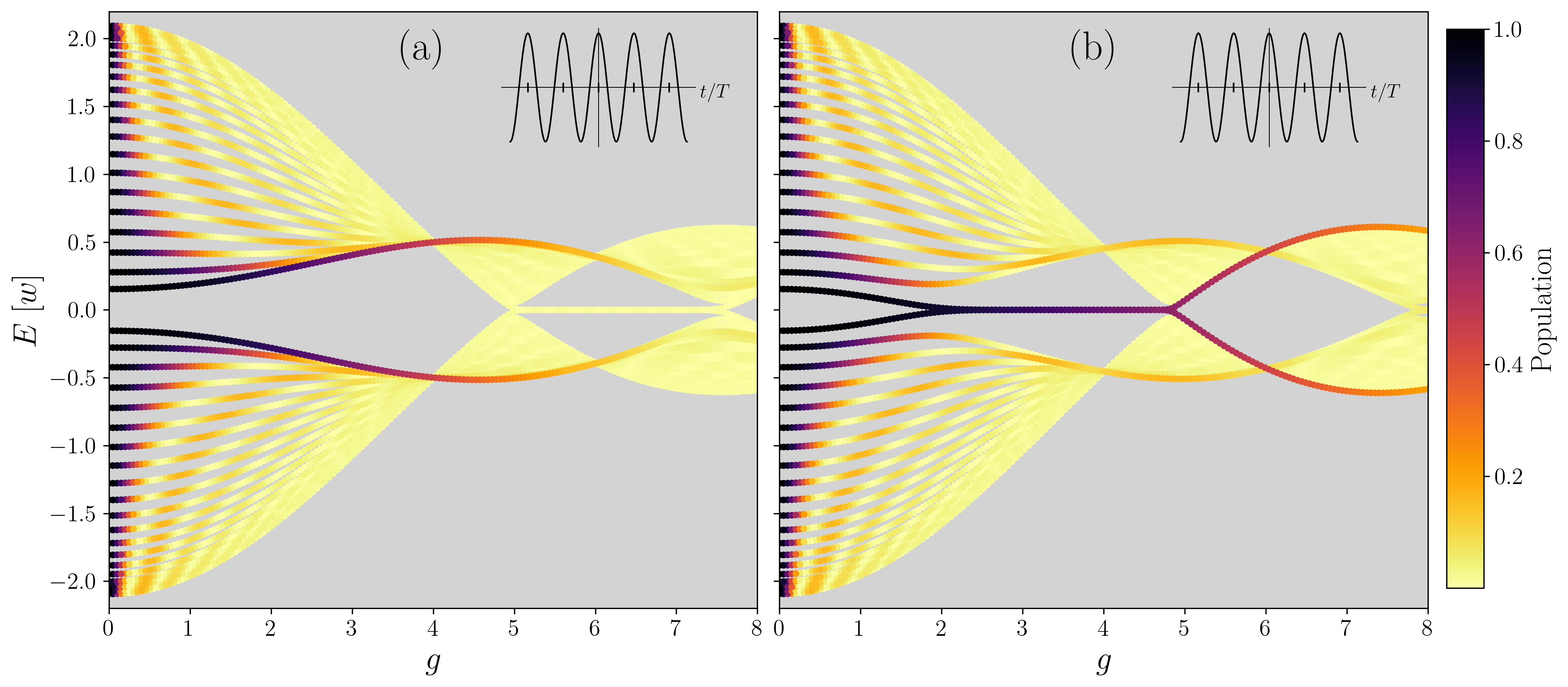}
\caption{Quasienergy spectrum within the off-resonant regime $\hbar\Omega/w=10$, for the static trivial phase $v/w=1.1$ with (a) $b=0.4a$ and (b)~$b/a = 0.6$ dimer separation.}
\label{fig:topological}
\end{figure*}

The next step is to confirm that, using the driving protocol, we can populate a topological state from a static, trivial phase of a SSH chain [see Figs.~\ref{fig:topological}(a) and~(b)]. Similarly to the behavior of the topological case shown in Fig.~\ref{fig:Trivial}, when $b < 0.5$ the first topological state is not populated. However, when $b>0.5$, the topological state is dominated by the two less energetic states, and so it is populated. This transition in the population of the first topological state came from the modulation rates of the hopping energies. 
When analyzing the continuum model of a static SSH chain, we find that the cases $w=0$, $v=\nu$ and $w=\nu$, $v=0$ exhibit identical quasienergy spectra since $E^2 = v^2 + w^2 + 2vw\cos q$. The only difference between them lies in the presence of surface states in the latter and their absence in the other. This spectral equivalence, despite the different topological characteristics, is reflected in Fig.~\ref{fig:topological} (a) and (b) around $g=4$.

The origin of this behavior lies in the parameter $b$. When $b/a<0.5$, the effective hopping amplitude that vanishes in that electric field is the one proportional to $w$, whereas for $b/a>0.5$, it is the amplitude proportional to $v$ that becomes zero. Furthermore, as shown in Eq.~\eqref{eq: HFloquet}, when $b/a$ approaches 1,  condition $v<w$ is satisfied at lower values of $g$, which means that the system enters the topological phase for smaller amplitudes of the driving field. This is most relevant since the electric field amplitude  is scaled by small quantities as the lattice parameter $a\sim1\, \rm{nm}$ and we are using high-frequency lasers. As a result, the actual electric field amplitude corresponds to $g\times10^{9}\, \rm{V/m}$ making it crucial to minimize $g$ to keep the required field strength experimentally feasible.
To conclude this section, it is worth to stress that while the inter-dimer separation instantaneously breaks the chiral symmetry when the ac electric field is applied, we have shown that there are some values of the amplitude where not only there is a topological phase, but its associated edge states are populated.

\section{Driving by electric field with modulated amplitude}\label{sec:driving}

We now extend the analysis to amplitude modulated driving protocols, where the amplitude of the driving field varies in time. To make connection to practical realizations, we focus on two relevant experimental scenarios: a Gaussian pulse, and a periodic envelope laser pulse. The standard approach to this problem relies on the $t-t'$ formalism~\cite{Drese1999, Holthaus2015, Ikeda2022, Baba_2025, Alejandro_2025}, which introduces two separate time scales. This allows the system to be treated as if it were driven instantaneously by a field with constant amplitude. 
However, one significant limitation of this method is its inapplicability in cases where the amplitude modulation is faster than the driving frequency as it assumes quasi-static driving conditions. For that reason, we will make use of the unitary transformation as defined in equation~\eqref{eq:transf_gen}. Up to equation~\eqref{eq:General}, we considered a general expression for the electric field. Therefore, let us make this equation our starting point for this section. 

First, we define the function $F(t) = \Omega \, \mathcal{E}_0 f(t) \xi(t)$ where $f(t)$ is the envelope function of the periodic driving $\xi(t)$ with frequency $\Omega$. Thus, the Hamiltonian \eqref{eq:H_transf}  can be written as
%
\begin{align}
    \mathcal{H}(t) &= v \sum_{\ell = 1}^N \ket{\ell}\bra{\ell}\Big[\exp\biggl(-ig^{\prime}\!\!\!\int F(t) dt\biggr)\sigma_+ + \mathrm{h.c.}\Big] \nonumber\\
    +& w \sum_{\ell = 1}^N \Big[\ket{\ell+1}\bra{\ell}\exp\biggl(ig^{\prime\prime}\!\!\!\int F(t) dt\biggr)\sigma_+ + \mathrm{h.c.}\Big]\ ,
    \label{eq:H_eff_env_mod}
\end{align}    
%
where $g^{\prime}$ and $g^{\prime\prime}$ are the same as defined before. Typically, the envelope function will not be periodic or it will have a different period than the driving. For that reason, the hopping modifiers cannot be easily decomposed into a discrete Floquet basis as it was in the static scenario, because now they are time-dependent. To cope with this problem, we will consider the coefficients of the decomposition as $a_n = \int_0^T a_n(t)dt$, there $T$ is the time period of the driving $T=2\pi/\Omega$ (or $\Omega_-$ in the periodic envelope laser pulse discussed below).

\begin{figure*}[ht] 
\centering
\includegraphics[width=0.8\linewidth]{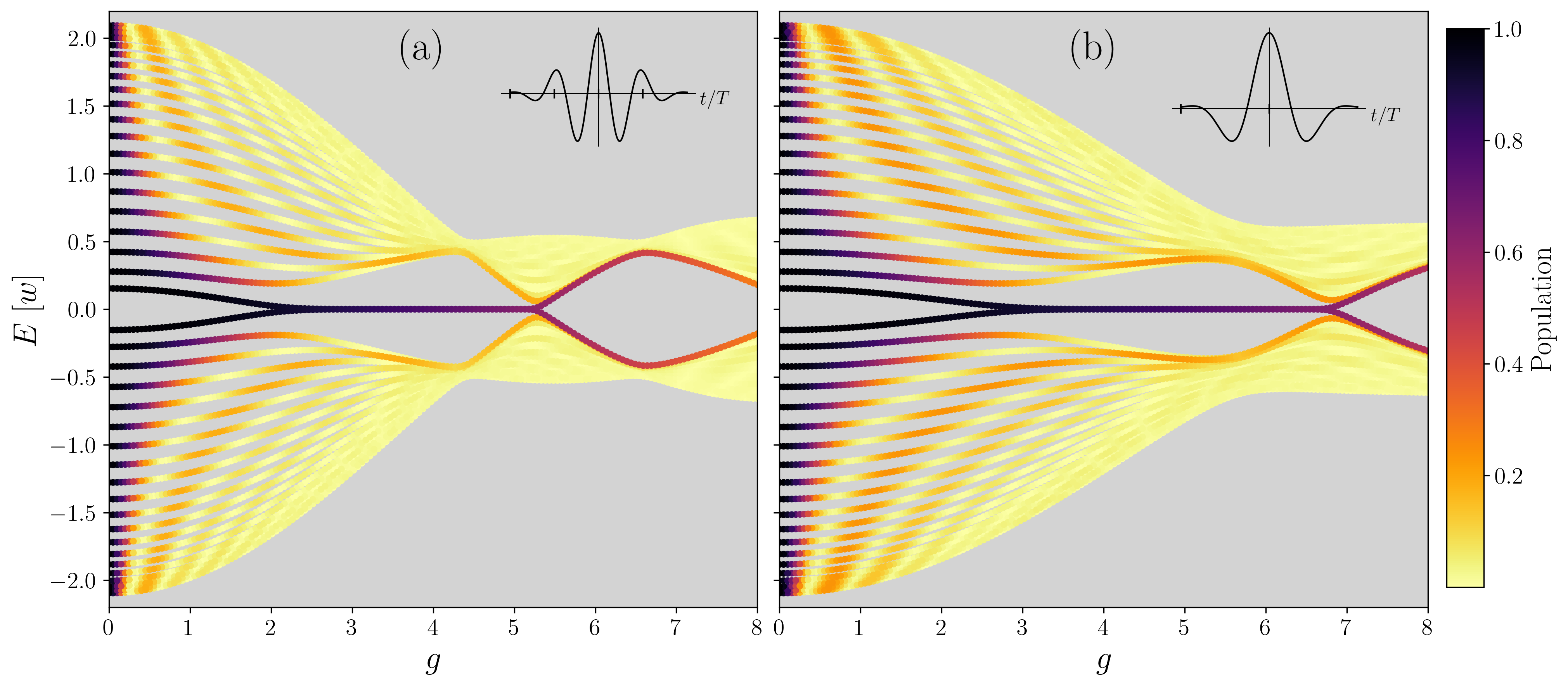}
\caption{Quasienergy spectrum within the off-resonant regime for the static trivial phase $v/w=1.1$ with inter-dimer separation $b/a = 0.6$ for a Gaussian pulse with (a)~$\Gamma=1$ s$^{-1}$ and (b)~$\Gamma=2$ s$^{-1}$. The color scale represents the population associated with each quasi-energy state. The frequency of the electric field has been set to $\hbar \Omega/w = 10$. The inset shows the time-dependent profile of the applied electric field, and the ticks in it's $x$-axis represent the integrating regions of Eq.~\eqref{Floquet-Fourier}.}
\label{fig:Gaussian}
\end{figure*}

\subsection{Gaussian pulse}

One of the most common pulse shapes used in driving experiments and measurements is the Gaussian envelope. Consequently, it is natural to investigate how the system behaves when the driving field is modulated by a Gaussian envelope function. we describe it as
\begin{equation}
\mathcal{E}(t)=\mathcal{E}_0e^{-(t\Gamma)^2}\cos\Omega t,
\end{equation}
with $\Gamma^{-1}$ being proportional to the width of the pulse. Introducing dimensionless variables $\tau=\Omega t$, $c=\Omega/\Gamma$, we get for the modulation factor
\begin{equation}
\int F(t) dt = \frac{\sqrt{\pi} c e^{-c^2/ 4}}{2} \,\Re \left(\text{erf}\left(  \frac{\tau}{c}+\frac{ic}{2}\right)\right)\ ,
\label{eq:Modulation_gauss}
\end{equation}
with the error function defined as
\begin{equation}
\text{erf}(z)=\frac{2}{\sqrt{\pi }}\int _0^z e^{-\xi^2}d \xi\ ,
\end{equation}
where $z$ is an arbitrary complex variable. As expected, in the limit $c\rightarrow\infty$, that is, $\Gamma\rightarrow 0$, we recover the former results for the constant-amplitude driving field.

Now, the next step is substituting Eq.~\eqref{eq:Modulation_gauss} into Eq.~\eqref{eq:H_eff_env_mod} and using the definition of the Floquet-Fourier Hamiltonian (Eq.~\eqref{Floquet-Fourier}) to find the quasienergy spectrum and populations. From Fig.~\ref{fig:Gaussian} we see that the effect of the Gaussian pulse is to lift the degeneracy at the points around $g=4$ and $g=6.5$ where one of the hopping parameters became zero. This lifting results in a spread of the quasienergy levels, keeping the lowest-energy states the most populated and thereby facilitating the occupation of subsequent topological states. However, we also observe that as the pulse becomes shorter, as in the case of Fig.  \ref{fig:Gaussian} (b) where $\Gamma = 5 \, \rm{s^{-1}}$, the spectra is stretched to higher $g$, delaying the emergence of the topological state.

\begin{figure*}[ht]
    \centering
    \includegraphics[width=0.8\linewidth]{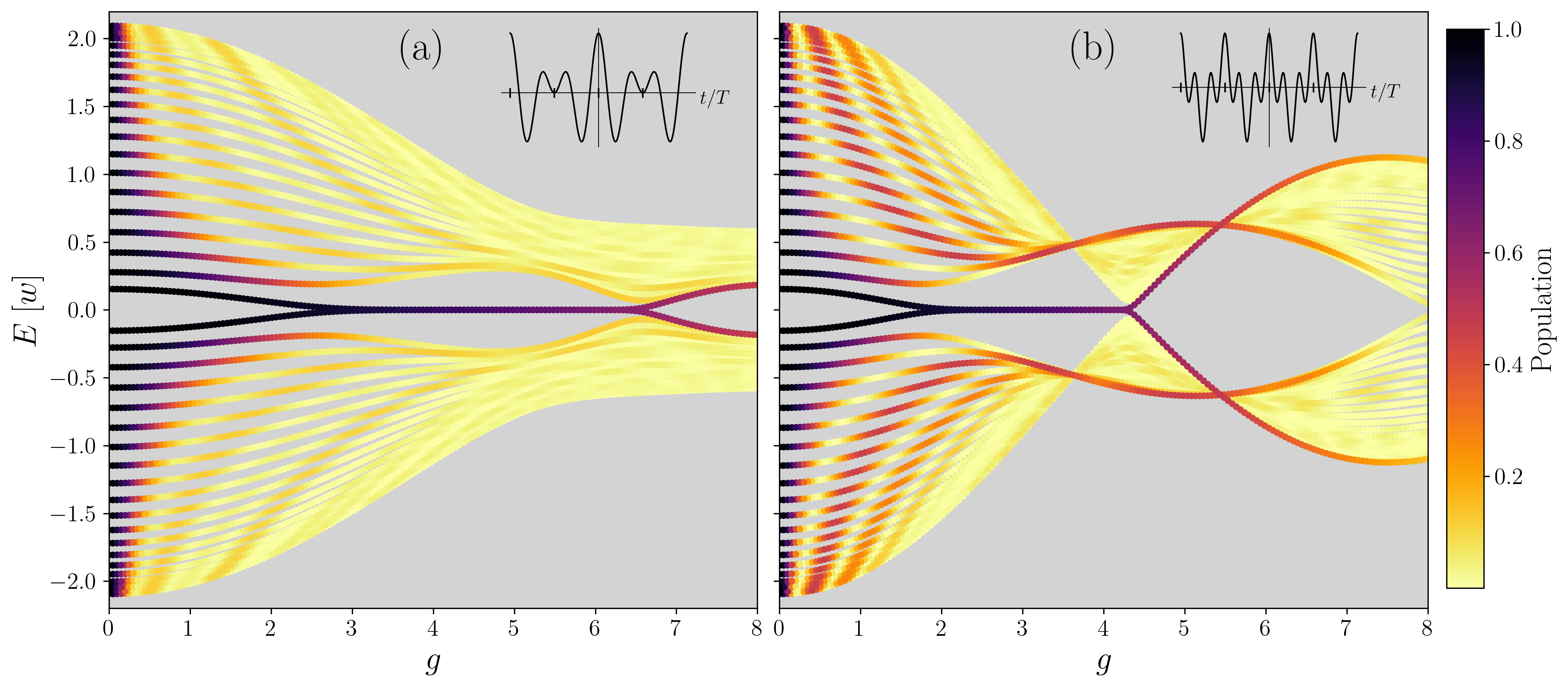}
    \caption{Quasienergy spectrum within the off-resonant regime for the static trivial phase $v/w=1.1$ with inter-dimer separation $b/a = 0.6$ for a periodic envelope laser pulse with (a)~$\hbar\omega/w=2$ and (b)~$\hbar\omega/w=5$. The color scale represents the population associated with each quasi-energy state. The frequency of the electric field has been set to $\hbar \Omega/w = 10$. The inset shows the time-dependent profile of the applied electric field, and the ticks in it's $x$-axis represent the integrating regions of Eq.~\eqref{Floquet-Fourier}.}
    \label{fig:Cosine}
\end{figure*}

\subsection{Periodic envelope laser pulse}

We now consider the driving-field configuration,
\begin{equation}
\mathcal{E}(t)=\mathcal{E}_0f(t)\cos\Omega t\ ,    
\end{equation}
where $\mathcal{E}$ is the maximum field amplitude, $\hbar\Omega \gg w,v$ and $f(t+T_2)=f(t)$, where $\omega=2\pi/T_2$ is the frequency of the periodically varying amplitude, which does not need to be commensurate with the driving frequency.
For instance, in the case $f(t)=\cos\omega t$ we get
\begin{equation}
\mathcal{E}(t)=\mathcal{E}_0\cos(\Omega t)\cos(\omega t),
\end{equation}
which corresponds to a superposition of two laser pulses of equal amplitude
producing a beating pattern of frequency $\Omega_-=\Omega-\omega$ and oscillatory  frequency $\Omega_+=\Omega+\omega$. Then, the effective electric field can be written as
\begin{equation}
\mathcal{E}(t)=\frac{\mathcal{E}_0}{2}(\cos\Omega_+ t+\cos\Omega_- t)\ .
\end{equation}
So we can understand this modulation as the use of two different harmonic waves for the driving, one for driving (fast one) and other for probe. The effective modulation of the hopping reads now
\begin{equation}
    \int F(t)dt = \frac{\Omega}{2\Omega_{+}}\sin \Omega_{+}t +\frac{\Omega}{2\Omega_{-}}\sin \Omega_{-}t\ ,
\end{equation}
%
where when $\omega\approx0$ we have $\Omega_+ \approx \Omega_- \approx \Omega$ so we restore the situation previously discussed.

Here, we can use the Jacobi-Anger expansion to write the effective intracell hopping modulation $p_v(t)$ as
\begin{align}
    p_v(t)&=\sum_{m_1 = -\infty}^{\infty}J_{m_1}\left(-g^{\prime}\,\frac{\Omega}{\Omega_+}\right) e^{im_1\Omega_+ t} \nonumber \\
    & \times \sum_{m_2= -\infty}^{\infty}J_{m_2}\left(-g^{\prime}\,\frac{\Omega}{\Omega_-}\right) e^{im_2\Omega_-t}\ ,
\label{eq:modulation_cosine}
\end{align}
with the effective modulation of the intercell hopping  $p_w(t)$ being the same with the change $g^{\prime}\rightarrow g^{\prime\prime}$. As driving is now composed of two distinct frequencies, the original driving frequency $\Omega$ is no longer representative of the system’s periodicity (see Eq.~\eqref{eq:modulation_cosine}). Therefore, in Eq.~\eqref{Floquet-Fourier}, we replace $\Omega$ by $\Omega_-$, the lower of the two driving frequencies. This choice is particularly relevant since selecting the lower frequency allows for a denser sampling of the quasi-energy spectrum, thereby capturing features that would be missed with a larger replica spacing such as that induced by $\Omega_+$.

The quasienergy spectra and the population are sketched in Fig.~\ref{fig:Cosine}. When the envelope frequency $\omega$ is small, the behavior resembles that of the Gaussian pulse: the degeneracies around specific values of $g$ are lifted, and the spectrum stretches towards higher field amplitudes. However, as $\omega$ increases and approaches the driving frequency $\Omega$, new degeneracies reappear in the spectrum and the transition to topological states changes to lower values of the effective light-matter coupling strength $g$.  This indicates a non-trivial interplay between the two time scales, which could be exploited to achieve these unconventional Floquet topological phases in the presence of dynamical chiral symmetry-breaking interactions. Notably, these phases can emerge under  less intense effective radiation fields compared to those required in the constant- amplitude driving protocol.

\section{Conclusions}\label{sec:conclusions}

In this work, we have analyzed the effect of an AC electric field applied to an SSH chain in the tuning of its topological phase. To this end, we employed a Krammers-Henneberger-like transformation, which effectively maps the contributions from all Floquet replicas onto the first one. 
We found that the system can be driven between the trivial and topological phases by varying the amplitude of the applied ac electric field. Moreover, we demonstrated that topological phases can persist even when the inter-dimer separation is nonzero and the chiral symmetry is explicitly broken due to time-periodic perturbation. To induce topological phases under realistic field strengths, larger inter-dimer separations are shown to be advantageous. 
It is worth noting that not all the topological states were populated in the topological phase. This is a crucial consideration, as the practical utility of these states---particularly for applications---relies on their actual population being finite.
Although nonperiodic driving protocols have traditionally been approached using approximations such as the $t-t'$ formalism, our Krammers-Henneberger-like transformation enables the treatment of arbitrary nonperiodic dynamical effects without imposing constraints on the rate of change of the perturbation.
We have also shown that Gaussian pulses shift the quasienergy spectrum to higher electric field amplitudes and lift degeneracies in the trivial energies, thereby enabling the population of higher-field topological states. In contrast, periodic envelope modulations can shift the spectrum to lower field amplitudes when the envelope frequency approaches the driving frequency ($\omega \sim \Omega$).
Therefore, our results suggest that the optimal configuration for experimental realization involves an SSH chain with large inter-dimer separation driven by a high-frequency electric field modulated with a fast periodic envelope. 

\begin{acknowledgments}

The authors thank G. Platero for helpful discussions. Work at Madrid has been supported by Comunidad de Madrid (Recovery, Transformation and Resilience Plan) and NextGenerationEU from the European Union (Grant MAD2D-CM-UCM5) and Agencia Estatal de Investigación (Grant PID2022-136285NB-C31).

\end{acknowledgments}

\bibliography{ssh_floquet}

\begin{thebibliography}{39}%
\makeatletter
\providecommand \@ifxundefined [1]{%
 \@ifx{#1\undefined}
}%
\providecommand \@ifnum [1]{%
 \ifnum #1\expandafter \@firstoftwo
 \else \expandafter \@secondoftwo
 \fi
}%
\providecommand \@ifx [1]{%
 \ifx #1\expandafter \@firstoftwo
 \else \expandafter \@secondoftwo
 \fi
}%
\providecommand \natexlab [1]{#1}%
\providecommand \enquote  [1]{``#1''}%
\providecommand \bibnamefont  [1]{#1}%
\providecommand \bibfnamefont [1]{#1}%
\providecommand \citenamefont [1]{#1}%
\providecommand \href@noop [0]{\@secondoftwo}%
\providecommand \href [0]{\begingroup \@sanitize@url \@href}%
\providecommand \@href[1]{\@@startlink{#1}\@@href}%
\providecommand \@@href[1]{\endgroup#1\@@endlink}%
\providecommand \@sanitize@url [0]{\catcode `\\12\catcode `\$12\catcode `\&12\catcode `\#12\catcode `\^12\catcode `\_12\catcode `\%12\relax}%
\providecommand \@@startlink[1]{}%
\providecommand \@@endlink[0]{}%
\providecommand \url  [0]{\begingroup\@sanitize@url \@url }%
\providecommand \@url [1]{\endgroup\@href {#1}{\urlprefix }}%
\providecommand \urlprefix  [0]{URL }%
\providecommand \Eprint [0]{\href }%
\providecommand \doibase [0]{https://doi.org/}%
\providecommand \selectlanguage [0]{\@gobble}%
\providecommand \bibinfo  [0]{\@secondoftwo}%
\providecommand \bibfield  [0]{\@secondoftwo}%
\providecommand \translation [1]{[#1]}%
\providecommand \BibitemOpen [0]{}%
\providecommand \bibitemStop [0]{}%
\providecommand \bibitemNoStop [0]{.\EOS\space}%
\providecommand \EOS [0]{\spacefactor3000\relax}%
\providecommand \BibitemShut  [1]{\csname bibitem#1\endcsname}%
\let\auto@bib@innerbib\@empty
\bibitem [{\citenamefont {{von Klitzing}}\ \emph {et~al.}(1980)\citenamefont {{von Klitzing}}, \citenamefont {Dorda},\ and\ \citenamefont {Pepper}}]{VonKlitzing1980}%
  \BibitemOpen
  \bibfield  {author} {\bibinfo {author} {\bibfnamefont {K.}~\bibnamefont {{von Klitzing}}}, \bibinfo {author} {\bibfnamefont {G.}~\bibnamefont {Dorda}},\ and\ \bibinfo {author} {\bibfnamefont {M.}~\bibnamefont {Pepper}},\ }\bibfield  {title} {\bibinfo {title} {New method for high-accuracy determination of the fine-structure constant based on quantized {H}all resistance},\ }\href {https://doi.org/10.1103/PhysRevLett.45.494} {\bibfield  {journal} {\bibinfo  {journal} {Phys. Rev. Lett.}\ }\textbf {\bibinfo {volume} {45}},\ \bibinfo {pages} {494} (\bibinfo {year} {1980})}\BibitemShut {NoStop}%
\bibitem [{\citenamefont {Haldane}(1988)}]{Haldane1988}%
  \BibitemOpen
  \bibfield  {author} {\bibinfo {author} {\bibfnamefont {F.~M.~D.}\ \bibnamefont {Haldane}},\ }\bibfield  {title} {\bibinfo {title} {Model for a quantum {H}all effect without {L}andau levels: Condensed-matter realization of the "parity anomaly"},\ }\href {https://doi.org/10.1103/PhysRevLett.61.2015} {\bibfield  {journal} {\bibinfo  {journal} {Phys. Rev. Lett.}\ }\textbf {\bibinfo {volume} {61}},\ \bibinfo {pages} {2015} (\bibinfo {year} {1988})}\BibitemShut {NoStop}%
\bibitem [{\citenamefont {Kane}\ and\ \citenamefont {Mele}(2005)}]{Kane2005}%
  \BibitemOpen
  \bibfield  {author} {\bibinfo {author} {\bibfnamefont {C.~L.}\ \bibnamefont {Kane}}\ and\ \bibinfo {author} {\bibfnamefont {E.~J.}\ \bibnamefont {Mele}},\ }\bibfield  {title} {\bibinfo {title} {Quantum spin {H}all effect in graphene},\ }\href {https://doi.org/10.1103/PhysRevLett.95.226801} {\bibfield  {journal} {\bibinfo  {journal} {Phys. Rev. Lett.}\ }\textbf {\bibinfo {volume} {95}},\ \bibinfo {pages} {226801} (\bibinfo {year} {2005})}\BibitemShut {NoStop}%
\bibitem [{\citenamefont {Bernevig}\ \emph {et~al.}(2006)\citenamefont {Bernevig}, \citenamefont {Hughes},\ and\ \citenamefont {Zhang}}]{Bernevig2006}%
  \BibitemOpen
  \bibfield  {author} {\bibinfo {author} {\bibfnamefont {B.~A.}\ \bibnamefont {Bernevig}}, \bibinfo {author} {\bibfnamefont {T.~L.}\ \bibnamefont {Hughes}},\ and\ \bibinfo {author} {\bibfnamefont {S.-C.}\ \bibnamefont {Zhang}},\ }\bibfield  {title} {\bibinfo {title} {Quantum spin hall effect and topological phase transition in hgte quantum wells},\ }\href {https://doi.org/10.1126/science.1133734} {\bibfield  {journal} {\bibinfo  {journal} {Science}\ }\textbf {\bibinfo {volume} {314}},\ \bibinfo {pages} {1757} (\bibinfo {year} {2006})}\BibitemShut {NoStop}%
\bibitem [{\citenamefont {K{\"o}nig}\ \emph {et~al.}(2007)\citenamefont {K{\"o}nig}, \citenamefont {Wiedmann}, \citenamefont {Br{\"u}ne}, \citenamefont {Roth}, \citenamefont {Buhmann}, \citenamefont {Molenkamp}, \citenamefont {Qi},\ and\ \citenamefont {Zhang}}]{Koenig2007}%
  \BibitemOpen
  \bibfield  {author} {\bibinfo {author} {\bibfnamefont {M.}~\bibnamefont {K{\"o}nig}}, \bibinfo {author} {\bibfnamefont {S.}~\bibnamefont {Wiedmann}}, \bibinfo {author} {\bibfnamefont {C.}~\bibnamefont {Br{\"u}ne}}, \bibinfo {author} {\bibfnamefont {A.}~\bibnamefont {Roth}}, \bibinfo {author} {\bibfnamefont {H.}~\bibnamefont {Buhmann}}, \bibinfo {author} {\bibfnamefont {L.~W.}\ \bibnamefont {Molenkamp}}, \bibinfo {author} {\bibfnamefont {X.-L.}\ \bibnamefont {Qi}},\ and\ \bibinfo {author} {\bibfnamefont {S.-C.}\ \bibnamefont {Zhang}},\ }\bibfield  {title} {\bibinfo {title} {Quantum spin {H}all insulator state in {HgTe} quantum wells},\ }\href {https://doi.org/10.1126/science.1148047} {\bibfield  {journal} {\bibinfo  {journal} {Science}\ }\textbf {\bibinfo {volume} {318}},\ \bibinfo {pages} {766} (\bibinfo {year} {2007})}\BibitemShut {NoStop}%
\bibitem [{\citenamefont {Hasan}\ and\ \citenamefont {Kane}(2010)}]{Hasan2010}%
  \BibitemOpen
  \bibfield  {author} {\bibinfo {author} {\bibfnamefont {M.~Z.}\ \bibnamefont {Hasan}}\ and\ \bibinfo {author} {\bibfnamefont {C.~L.}\ \bibnamefont {Kane}},\ }\bibfield  {title} {\bibinfo {title} {Colloquium: Topological insulators},\ }\href {https://doi.org/10.1103/RevModPhys.82.3045} {\bibfield  {journal} {\bibinfo  {journal} {Rev. Mod. Phys.}\ }\textbf {\bibinfo {volume} {82}},\ \bibinfo {pages} {3045} (\bibinfo {year} {2010})}\BibitemShut {NoStop}%
\bibitem [{\citenamefont {Alexandradinata}\ \emph {et~al.}(2014)\citenamefont {Alexandradinata}, \citenamefont {Fang}, \citenamefont {Gilbert},\ and\ \citenamefont {Bernevig}}]{Bernevig2014}%
  \BibitemOpen
  \bibfield  {author} {\bibinfo {author} {\bibfnamefont {A.}~\bibnamefont {Alexandradinata}}, \bibinfo {author} {\bibfnamefont {C.}~\bibnamefont {Fang}}, \bibinfo {author} {\bibfnamefont {M.~J.}\ \bibnamefont {Gilbert}},\ and\ \bibinfo {author} {\bibfnamefont {B.~A.}\ \bibnamefont {Bernevig}},\ }\bibfield  {title} {\bibinfo {title} {Spin-orbit-free topological insulators without time-reversal symmetry},\ }\href {https://doi.org/10.1103/PhysRevLett.113.116403} {\bibfield  {journal} {\bibinfo  {journal} {Phys. Rev. Lett.}\ }\textbf {\bibinfo {volume} {113}},\ \bibinfo {pages} {116403} (\bibinfo {year} {2014})}\BibitemShut {NoStop}%
\bibitem [{\citenamefont {Oka}\ and\ \citenamefont {Aoki}(2009)}]{Oka09}%
  \BibitemOpen
  \bibfield  {author} {\bibinfo {author} {\bibfnamefont {T.}~\bibnamefont {Oka}}\ and\ \bibinfo {author} {\bibfnamefont {H.}~\bibnamefont {Aoki}},\ }\bibfield  {title} {\bibinfo {title} {Photovoltaic {H}all effect in graphene},\ }\href {https://doi.org/10.1103/PhysRevB.79.081406} {\bibfield  {journal} {\bibinfo  {journal} {Phys. Rev. B}\ }\textbf {\bibinfo {volume} {79}},\ \bibinfo {pages} {081406} (\bibinfo {year} {2009})}\BibitemShut {NoStop}%
\bibitem [{\citenamefont {D\'ora}\ \emph {et~al.}(2009)\citenamefont {D\'ora}, \citenamefont {Ziegler}, \citenamefont {Thalmeier},\ and\ \citenamefont {Nakamura}}]{Dora2009}%
  \BibitemOpen
  \bibfield  {author} {\bibinfo {author} {\bibfnamefont {B.}~\bibnamefont {D\'ora}}, \bibinfo {author} {\bibfnamefont {K.}~\bibnamefont {Ziegler}}, \bibinfo {author} {\bibfnamefont {P.}~\bibnamefont {Thalmeier}},\ and\ \bibinfo {author} {\bibfnamefont {M.}~\bibnamefont {Nakamura}},\ }\bibfield  {title} {\bibinfo {title} {Rabi oscillations in {L}andau-quantized graphene},\ }\href {https://doi.org/10.1103/PhysRevLett.102.036803} {\bibfield  {journal} {\bibinfo  {journal} {Phys. Rev. Lett.}\ }\textbf {\bibinfo {volume} {102}},\ \bibinfo {pages} {036803} (\bibinfo {year} {2009})}\BibitemShut {NoStop}%
\bibitem [{\citenamefont {Lindner}\ \emph {et~al.}(2011)\citenamefont {Lindner}, \citenamefont {Refael},\ and\ \citenamefont {Galitski}}]{Lindner2011}%
  \BibitemOpen
  \bibfield  {author} {\bibinfo {author} {\bibfnamefont {N.~H.}\ \bibnamefont {Lindner}}, \bibinfo {author} {\bibfnamefont {G.}~\bibnamefont {Refael}},\ and\ \bibinfo {author} {\bibfnamefont {V.}~\bibnamefont {Galitski}},\ }\bibfield  {title} {\bibinfo {title} {Floquet topological insulator in semiconductor quantum wells},\ }\href {https://doi.org/10.1038/nphys1926} {\bibfield  {journal} {\bibinfo  {journal} {Nat. Phys.}\ }\textbf {\bibinfo {volume} {6}},\ \bibinfo {pages} {1745} (\bibinfo {year} {2011})}\BibitemShut {NoStop}%
\bibitem [{\citenamefont {Usaj}\ \emph {et~al.}(2014)\citenamefont {Usaj}, \citenamefont {Perez-Piskunow}, \citenamefont {Torres},\ and\ \citenamefont {Balseiro}}]{usaj2014}%
  \BibitemOpen
  \bibfield  {author} {\bibinfo {author} {\bibfnamefont {G.}~\bibnamefont {Usaj}}, \bibinfo {author} {\bibfnamefont {P.~M.}\ \bibnamefont {Perez-Piskunow}}, \bibinfo {author} {\bibfnamefont {L.~E. F.~F.}\ \bibnamefont {Torres}},\ and\ \bibinfo {author} {\bibfnamefont {C.~A.}\ \bibnamefont {Balseiro}},\ }\bibfield  {title} {\bibinfo {title} {Irradiated graphene as a tunable {F}loquet topological insulator},\ }\href {https://doi.org/10.1103/PhysRevB.90.115423} {\bibfield  {journal} {\bibinfo  {journal} {Phys. Rev. B}\ }\textbf {\bibinfo {volume} {90}},\ \bibinfo {pages} {115423} (\bibinfo {year} {2014})}\BibitemShut {NoStop}%
\bibitem [{\citenamefont {Grushin}\ \emph {et~al.}(2014)\citenamefont {Grushin}, \citenamefont {G\'omez-Le\'on},\ and\ \citenamefont {Neupert}}]{Grushin2014}%
  \BibitemOpen
  \bibfield  {author} {\bibinfo {author} {\bibfnamefont {A.~G.}\ \bibnamefont {Grushin}}, \bibinfo {author} {\bibfnamefont {A.}~\bibnamefont {G\'omez-Le\'on}},\ and\ \bibinfo {author} {\bibfnamefont {T.}~\bibnamefont {Neupert}},\ }\bibfield  {title} {\bibinfo {title} {Floquet fractional {C}hern insulators},\ }\href {https://doi.org/10.1103/PhysRevLett.112.156801} {\bibfield  {journal} {\bibinfo  {journal} {Phys. Rev. Lett.}\ }\textbf {\bibinfo {volume} {112}},\ \bibinfo {pages} {156801} (\bibinfo {year} {2014})}\BibitemShut {NoStop}%
\bibitem [{\citenamefont {Titum}\ \emph {et~al.}(2017)\citenamefont {Titum}, \citenamefont {Lindner},\ and\ \citenamefont {Refael}}]{Titum2017}%
  \BibitemOpen
  \bibfield  {author} {\bibinfo {author} {\bibfnamefont {P.}~\bibnamefont {Titum}}, \bibinfo {author} {\bibfnamefont {N.~H.}\ \bibnamefont {Lindner}},\ and\ \bibinfo {author} {\bibfnamefont {G.}~\bibnamefont {Refael}},\ }\bibfield  {title} {\bibinfo {title} {Disorder-induced transitions in resonantly driven {F}loquet topological insulators},\ }\href {https://doi.org/10.1103/PhysRevB.96.054207} {\bibfield  {journal} {\bibinfo  {journal} {Phys. Rev. B}\ }\textbf {\bibinfo {volume} {96}},\ \bibinfo {pages} {054207} (\bibinfo {year} {2017})}\BibitemShut {NoStop}%
\bibitem [{\citenamefont {Esin}\ \emph {et~al.}(2018)\citenamefont {Esin}, \citenamefont {Rudner}, \citenamefont {Refael},\ and\ \citenamefont {Netanel}}]{Esin2018}%
  \BibitemOpen
  \bibfield  {author} {\bibinfo {author} {\bibfnamefont {I.}~\bibnamefont {Esin}}, \bibinfo {author} {\bibfnamefont {M.~S.}\ \bibnamefont {Rudner}}, \bibinfo {author} {\bibfnamefont {G.}~\bibnamefont {Refael}},\ and\ \bibinfo {author} {\bibfnamefont {H.}~\bibnamefont {Netanel}},\ }\bibfield  {title} {\bibinfo {title} {Quantized transport and steady states of {F}loquet topological insulators},\ }\href {https://doi.org/10.1103/PhysRevB.97.245401} {\bibfield  {journal} {\bibinfo  {journal} {Phys. Rev. B}\ }\textbf {\bibinfo {volume} {97}},\ \bibinfo {pages} {245401} (\bibinfo {year} {2018})}\BibitemShut {NoStop}%
\bibitem [{\citenamefont {Peng}\ and\ \citenamefont {Refael}(2019)}]{Peng2019}%
  \BibitemOpen
  \bibfield  {author} {\bibinfo {author} {\bibfnamefont {Y.}~\bibnamefont {Peng}}\ and\ \bibinfo {author} {\bibfnamefont {G.}~\bibnamefont {Refael}},\ }\bibfield  {title} {\bibinfo {title} {Floquet second-order topological insulators from nonsymmorphic space-time symmetries},\ }\href {https://doi.org/10.1103/PhysRevLett.123.016806} {\bibfield  {journal} {\bibinfo  {journal} {Phys. Rev. Lett.}\ }\textbf {\bibinfo {volume} {123}},\ \bibinfo {pages} {016806} (\bibinfo {year} {2019})}\BibitemShut {NoStop}%
\bibitem [{\citenamefont {L\"u}\ and\ \citenamefont {Xie}(2019)}]{Lu2019}%
  \BibitemOpen
  \bibfield  {author} {\bibinfo {author} {\bibfnamefont {X.-L.}\ \bibnamefont {L\"u}}\ and\ \bibinfo {author} {\bibfnamefont {H.}~\bibnamefont {Xie}},\ }\bibfield  {title} {\bibinfo {title} {Topological phases and pumps in the {Su-Schrieffer-Heeger} model periodically modulated in time},\ }\href {https://doi.org/10.1088/1361-648X/ab3d72} {\bibfield  {journal} {\bibinfo  {journal} {J. Phys. Condens. Matter}\ }\textbf {\bibinfo {volume} {31}},\ \bibinfo {pages} {495401} (\bibinfo {year} {2019})}\BibitemShut {NoStop}%
\bibitem [{\citenamefont {Kitaev}(2009)}]{Kitaev2009}%
  \BibitemOpen
  \bibfield  {author} {\bibinfo {author} {\bibfnamefont {A.}~\bibnamefont {Kitaev}},\ }\bibfield  {title} {\bibinfo {title} {Periodic table for topological insulators and superconductors},\ }\href {https://doi.org/10.48550/arXiv.0901.2686} {\bibfield  {journal} {\bibinfo  {journal} {AIP Conf. Proc.}\ }\textbf {\bibinfo {volume} {1134}},\ \bibinfo {pages} {22} (\bibinfo {year} {2009})}\BibitemShut {NoStop}%
\bibitem [{\citenamefont {Schnyder}\ \emph {et~al.}(2008)\citenamefont {Schnyder}, \citenamefont {Ryu}, \citenamefont {Furusaki},\ and\ \citenamefont {Ludwig}}]{Schnyder2008}%
  \BibitemOpen
  \bibfield  {author} {\bibinfo {author} {\bibfnamefont {A.~P.}\ \bibnamefont {Schnyder}}, \bibinfo {author} {\bibfnamefont {S.}~\bibnamefont {Ryu}}, \bibinfo {author} {\bibfnamefont {A.}~\bibnamefont {Furusaki}},\ and\ \bibinfo {author} {\bibfnamefont {A.~W.~W.}\ \bibnamefont {Ludwig}},\ }\bibfield  {title} {\bibinfo {title} {Classification of topological insulators and superconductors in three spatial dimensions},\ }\href {https://doi.org/10.1103/PhysRevB.78.195125} {\bibfield  {journal} {\bibinfo  {journal} {Phys. Rev. B}\ }\textbf {\bibinfo {volume} {78}},\ \bibinfo {pages} {195125} (\bibinfo {year} {2008})}\BibitemShut {NoStop}%
\bibitem [{\citenamefont {Chiu}\ \emph {et~al.}(2016)\citenamefont {Chiu}, \citenamefont {Teo}, \citenamefont {Schnyder},\ and\ \citenamefont {Ryu}}]{Chiu2016}%
  \BibitemOpen
  \bibfield  {author} {\bibinfo {author} {\bibfnamefont {C.-K.}\ \bibnamefont {Chiu}}, \bibinfo {author} {\bibfnamefont {J.~C.~Y.}\ \bibnamefont {Teo}}, \bibinfo {author} {\bibfnamefont {A.~P.}\ \bibnamefont {Schnyder}},\ and\ \bibinfo {author} {\bibfnamefont {S.}~\bibnamefont {Ryu}},\ }\bibfield  {title} {\bibinfo {title} {Classification of topological quantum matter with symmetries},\ }\href {https://doi.org/10.1103/RevModPhys.88.035005} {\bibfield  {journal} {\bibinfo  {journal} {Rev. Mod. Phys.}\ }\textbf {\bibinfo {volume} {88}},\ \bibinfo {pages} {035005} (\bibinfo {year} {2016})}\BibitemShut {NoStop}%
\bibitem [{\citenamefont {Chiu}\ \emph {et~al.}(2013)\citenamefont {Chiu}, \citenamefont {Yao},\ and\ \citenamefont {Ryu}}]{Chiu2013}%
  \BibitemOpen
  \bibfield  {author} {\bibinfo {author} {\bibfnamefont {C.-K.}\ \bibnamefont {Chiu}}, \bibinfo {author} {\bibfnamefont {H.}~\bibnamefont {Yao}},\ and\ \bibinfo {author} {\bibfnamefont {S.}~\bibnamefont {Ryu}},\ }\bibfield  {title} {\bibinfo {title} {Classification of topological insulators and superconductors in the presence of reflection symmetry},\ }\href {https://doi.org/10.1103/PhysRevB.88.075142} {\bibfield  {journal} {\bibinfo  {journal} {Phys. Rev. B}\ }\textbf {\bibinfo {volume} {88}},\ \bibinfo {pages} {075142} (\bibinfo {year} {2013})}\BibitemShut {NoStop}%
\bibitem [{\citenamefont {Langbehn}\ \emph {et~al.}(2017)\citenamefont {Langbehn}, \citenamefont {Peng}, \citenamefont {Trifunovic}, \citenamefont {von Oppen},\ and\ \citenamefont {Brouwer}}]{Langbehn2017}%
  \BibitemOpen
  \bibfield  {author} {\bibinfo {author} {\bibfnamefont {J.}~\bibnamefont {Langbehn}}, \bibinfo {author} {\bibfnamefont {Y.}~\bibnamefont {Peng}}, \bibinfo {author} {\bibfnamefont {L.}~\bibnamefont {Trifunovic}}, \bibinfo {author} {\bibfnamefont {F.}~\bibnamefont {von Oppen}},\ and\ \bibinfo {author} {\bibfnamefont {P.~W.}\ \bibnamefont {Brouwer}},\ }\bibfield  {title} {\bibinfo {title} {Reflection-symmetric second-order topological insulators and superconductors},\ }\href {https://doi.org/10.1103/PhysRevLett.119.246401} {\bibfield  {journal} {\bibinfo  {journal} {Phys. Rev. Lett.}\ }\textbf {\bibinfo {volume} {119}},\ \bibinfo {pages} {246401} (\bibinfo {year} {2017})}\BibitemShut {NoStop}%
\bibitem [{\citenamefont {G\'omez-Le\'on}\ and\ \citenamefont {Platero}(2013)}]{Platero2013}%
  \BibitemOpen
  \bibfield  {author} {\bibinfo {author} {\bibfnamefont {A.}~\bibnamefont {G\'omez-Le\'on}}\ and\ \bibinfo {author} {\bibfnamefont {G.}~\bibnamefont {Platero}},\ }\bibfield  {title} {\bibinfo {title} {Floquet-bloch theory and topology in periodically driven lattices},\ }\href {https://doi.org/10.1103/PhysRevLett.110.200403} {\bibfield  {journal} {\bibinfo  {journal} {Phys. Rev. Lett.}\ }\textbf {\bibinfo {volume} {110}},\ \bibinfo {pages} {200403} (\bibinfo {year} {2013})}\BibitemShut {NoStop}%
\bibitem [{\citenamefont {Baba}\ \emph {et~al.}(2025)\citenamefont {Baba}, \citenamefont {Junk}, \citenamefont {Hogger}, \citenamefont {Domínguez-Adame}, \citenamefont {Molina},\ and\ \citenamefont {Richter}}]{Baba_2025}%
  \BibitemOpen
  \bibfield  {author} {\bibinfo {author} {\bibfnamefont {Y.}~\bibnamefont {Baba}}, \bibinfo {author} {\bibfnamefont {V.}~\bibnamefont {Junk}}, \bibinfo {author} {\bibfnamefont {W.}~\bibnamefont {Hogger}}, \bibinfo {author} {\bibfnamefont {F.}~\bibnamefont {Domínguez-Adame}}, \bibinfo {author} {\bibfnamefont {R.~A.}\ \bibnamefont {Molina}},\ and\ \bibinfo {author} {\bibfnamefont {K.}~\bibnamefont {Richter}},\ }\bibfield  {title} {\bibinfo {title} {Radiation-induced dynamical formation of {F}loquet-{B}loch bands in {D}irac {H}amiltonians},\ }\href {https://doi.org/10.1088/1367-2630/adc594} {\bibfield  {journal} {\bibinfo  {journal} {New J. Phys.}\ }\textbf {\bibinfo {volume} {27}},\ \bibinfo {pages} {043015} (\bibinfo {year} {2025})}\BibitemShut {NoStop}%
\bibitem [{\citenamefont {Ito}\ \emph {et~al.}(2023)\citenamefont {Ito}, \citenamefont {Schüler}, \citenamefont {Meierhofer}, \citenamefont {Schlauderer}, \citenamefont {Freudenstein}, \citenamefont {Reimann}, \citenamefont {Afanasiev}, \citenamefont {Kokh}, \citenamefont {Tereshchenko}, \citenamefont {Güdde}, \citenamefont {Sentef}, \citenamefont {Höfer},\ and\ \citenamefont {Huber}}]{Ito2023}%
  \BibitemOpen
  \bibfield  {author} {\bibinfo {author} {\bibfnamefont {S.}~\bibnamefont {Ito}}, \bibinfo {author} {\bibfnamefont {M.}~\bibnamefont {Schüler}}, \bibinfo {author} {\bibfnamefont {M.}~\bibnamefont {Meierhofer}}, \bibinfo {author} {\bibfnamefont {S.}~\bibnamefont {Schlauderer}}, \bibinfo {author} {\bibfnamefont {J.}~\bibnamefont {Freudenstein}}, \bibinfo {author} {\bibfnamefont {J.}~\bibnamefont {Reimann}}, \bibinfo {author} {\bibfnamefont {D.}~\bibnamefont {Afanasiev}}, \bibinfo {author} {\bibfnamefont {K.~A.}\ \bibnamefont {Kokh}}, \bibinfo {author} {\bibfnamefont {O.~E.}\ \bibnamefont {Tereshchenko}}, \bibinfo {author} {\bibfnamefont {J.}~\bibnamefont {Güdde}}, \bibinfo {author} {\bibfnamefont {M.~A.}\ \bibnamefont {Sentef}}, \bibinfo {author} {\bibfnamefont {U.}~\bibnamefont {Höfer}},\ and\ \bibinfo {author} {\bibfnamefont {R.}~\bibnamefont {Huber}},\ }\bibfield  {title} {\bibinfo {title} {Build-up and dephasing of {F}loquet-{B}loch bands on subcycle timescales},\ }\href
  {https://doi.org/10.1038/s41586-023-05850-x} {\bibfield  {journal} {\bibinfo  {journal} {Nature}\ }\textbf {\bibinfo {volume} {616}},\ \bibinfo {pages} {696} (\bibinfo {year} {2023})}\BibitemShut {NoStop}%
\bibitem [{\citenamefont {McIver}\ \emph {et~al.}(2019)\citenamefont {McIver}, \citenamefont {Schulte}, \citenamefont {Stein}, \citenamefont {Matsuyama}, \citenamefont {Jotzu}, \citenamefont {Meier},\ and\ \citenamefont {Cavalleri}}]{McIver2019}%
  \BibitemOpen
  \bibfield  {author} {\bibinfo {author} {\bibfnamefont {J.~W.}\ \bibnamefont {McIver}}, \bibinfo {author} {\bibfnamefont {B.}~\bibnamefont {Schulte}}, \bibinfo {author} {\bibfnamefont {F.-U.}\ \bibnamefont {Stein}}, \bibinfo {author} {\bibfnamefont {T.}~\bibnamefont {Matsuyama}}, \bibinfo {author} {\bibfnamefont {G.}~\bibnamefont {Jotzu}}, \bibinfo {author} {\bibfnamefont {G.}~\bibnamefont {Meier}},\ and\ \bibinfo {author} {\bibfnamefont {A.}~\bibnamefont {Cavalleri}},\ }\bibfield  {title} {\bibinfo {title} {Light-induced anomalous {H}all effect in graphene},\ }\href {https://doi.org/10.1038/s41567-019-0698-y} {\bibfield  {journal} {\bibinfo  {journal} {Nat. Phys.}\ }\textbf {\bibinfo {volume} {16}},\ \bibinfo {pages} {38} (\bibinfo {year} {2019})}\BibitemShut {NoStop}%
\bibitem [{\citenamefont {Merboldt}\ \emph {et~al.}(2025)\citenamefont {Merboldt}, \citenamefont {Schüler}, \citenamefont {Schmitt}, \citenamefont {Bange}, \citenamefont {Bennecke}, \citenamefont {Gadge}, \citenamefont {Pierz}, \citenamefont {Schumacher}, \citenamefont {Momeni}, \citenamefont {Steil}, \citenamefont {Manmana}, \citenamefont {Sentef}, \citenamefont {Reutzel},\ and\ \citenamefont {Mathias}}]{Merboldt2025}%
  \BibitemOpen
  \bibfield  {author} {\bibinfo {author} {\bibfnamefont {M.}~\bibnamefont {Merboldt}}, \bibinfo {author} {\bibfnamefont {M.}~\bibnamefont {Schüler}}, \bibinfo {author} {\bibfnamefont {D.}~\bibnamefont {Schmitt}}, \bibinfo {author} {\bibfnamefont {J.~P.}\ \bibnamefont {Bange}}, \bibinfo {author} {\bibfnamefont {W.}~\bibnamefont {Bennecke}}, \bibinfo {author} {\bibfnamefont {K.}~\bibnamefont {Gadge}}, \bibinfo {author} {\bibfnamefont {K.}~\bibnamefont {Pierz}}, \bibinfo {author} {\bibfnamefont {H.~W.}\ \bibnamefont {Schumacher}}, \bibinfo {author} {\bibfnamefont {D.}~\bibnamefont {Momeni}}, \bibinfo {author} {\bibfnamefont {D.}~\bibnamefont {Steil}}, \bibinfo {author} {\bibfnamefont {S.~R.}\ \bibnamefont {Manmana}}, \bibinfo {author} {\bibfnamefont {M.~A.}\ \bibnamefont {Sentef}}, \bibinfo {author} {\bibfnamefont {M.}~\bibnamefont {Reutzel}},\ and\ \bibinfo {author} {\bibfnamefont {S.}~\bibnamefont {Mathias}},\ }\bibfield  {title} {\bibinfo {title} {Observation of {F}loquet states in graphene},\ }\href
  {https://doi.org/10.1038/s41567-025-02889-7} {\bibfield  {journal} {\bibinfo  {journal} {Nat. Phys.}\ }\textbf {\bibinfo {volume} {21}},\ \bibinfo {pages} {1093} (\bibinfo {year} {2025})}\BibitemShut {NoStop}%
\bibitem [{\citenamefont {Choi}\ \emph {et~al.}(2025)\citenamefont {Choi}, \citenamefont {Mogi}, \citenamefont {De~Giovannini}, \citenamefont {Azoury}, \citenamefont {Lv}, \citenamefont {Su}, \citenamefont {Hübener}, \citenamefont {Rubio},\ and\ \citenamefont {Gedik}}]{Choi2025}%
  \BibitemOpen
  \bibfield  {author} {\bibinfo {author} {\bibfnamefont {D.}~\bibnamefont {Choi}}, \bibinfo {author} {\bibfnamefont {M.}~\bibnamefont {Mogi}}, \bibinfo {author} {\bibfnamefont {U.}~\bibnamefont {De~Giovannini}}, \bibinfo {author} {\bibfnamefont {D.}~\bibnamefont {Azoury}}, \bibinfo {author} {\bibfnamefont {B.}~\bibnamefont {Lv}}, \bibinfo {author} {\bibfnamefont {Y.}~\bibnamefont {Su}}, \bibinfo {author} {\bibfnamefont {H.}~\bibnamefont {Hübener}}, \bibinfo {author} {\bibfnamefont {A.}~\bibnamefont {Rubio}},\ and\ \bibinfo {author} {\bibfnamefont {N.}~\bibnamefont {Gedik}},\ }\bibfield  {title} {\bibinfo {title} {Observation of {F}loquet–{B}loch states in monolayer graphene},\ }\href {https://doi.org/10.1038/s41567-025-02888-8} {\bibfield  {journal} {\bibinfo  {journal} {Nat. Phys.}\ }\textbf {\bibinfo {volume} {21}},\ \bibinfo {pages} {1093} (\bibinfo {year} {2025})}\BibitemShut {NoStop}%
\bibitem [{\citenamefont {Drese}\ and\ \citenamefont {Holthaus}(1999)}]{Drese1999}%
  \BibitemOpen
  \bibfield  {author} {\bibinfo {author} {\bibfnamefont {K.}~\bibnamefont {Drese}}\ and\ \bibinfo {author} {\bibfnamefont {M.}~\bibnamefont {Holthaus}},\ }\bibfield  {title} {\bibinfo {title} {Floquet theory for short laser pulses},\ }\href {https://doi.org/10.1007/s100530050236} {\bibfield  {journal} {\bibinfo  {journal} {Eur. Phys. J. D}\ }\textbf {\bibinfo {volume} {5}},\ \bibinfo {pages} {119} (\bibinfo {year} {1999})}\BibitemShut {NoStop}%
\bibitem [{\citenamefont {Holthaus}(2015)}]{Holthaus2015}%
  \BibitemOpen
  \bibfield  {author} {\bibinfo {author} {\bibfnamefont {M.}~\bibnamefont {Holthaus}},\ }\bibfield  {title} {\bibinfo {title} {Floquet engineering with quasienergy bands of periodically driven optical lattices},\ }\href {https://doi.org/10.1088/0953-4075/49/1/013001} {\bibfield  {journal} {\bibinfo  {journal} {Eur. Phys. J. B}\ }\textbf {\bibinfo {volume} {49}},\ \bibinfo {pages} {013001} (\bibinfo {year} {2015})}\BibitemShut {NoStop}%
\bibitem [{\citenamefont {Ikeda}\ \emph {et~al.}(2022)\citenamefont {Ikeda}, \citenamefont {Tanaka},\ and\ \citenamefont {Kayanuma}}]{Ikeda2022}%
  \BibitemOpen
  \bibfield  {author} {\bibinfo {author} {\bibfnamefont {T.~N.}\ \bibnamefont {Ikeda}}, \bibinfo {author} {\bibfnamefont {S.}~\bibnamefont {Tanaka}},\ and\ \bibinfo {author} {\bibfnamefont {Y.}~\bibnamefont {Kayanuma}},\ }\bibfield  {title} {\bibinfo {title} {Floquet-{L}andau-{Z}ener interferometry: Usefulness of the floquet theory in pulse-laser-driven systems},\ }\href {https://doi.org/10.1103/physrevresearch.4.033075} {\bibfield  {journal} {\bibinfo  {journal} {Phys. Rev. Res.}\ }\textbf {\bibinfo {volume} {4}},\ \bibinfo {pages} {033075} (\bibinfo {year} {2022})}\BibitemShut {NoStop}%
\bibitem [{\citenamefont {G\'omez}\ \emph {et~al.}(2025)\citenamefont {G\'omez}, \citenamefont {Baba}, \citenamefont {Dom\'inguez-Adame},\ and\ \citenamefont {Molina}}]{Alejandro_2025}%
  \BibitemOpen
  \bibfield  {author} {\bibinfo {author} {\bibfnamefont {A.~S.}\ \bibnamefont {G\'omez}}, \bibinfo {author} {\bibfnamefont {Y.}~\bibnamefont {Baba}}, \bibinfo {author} {\bibfnamefont {F.}~\bibnamefont {Dom\'inguez-Adame}},\ and\ \bibinfo {author} {\bibfnamefont {R.~A.}\ \bibnamefont {Molina}},\ }\bibfield  {title} {\bibinfo {title} {Time-dependent dichroism and transient valley polarization in monolayer transition metal dichalcogenides under finite-pulse radiation},\ }\href {https://doi.org/10.1088/2515-7639/ade292} {\bibfield  {journal} {\bibinfo  {journal} {J. Phys. Mater.}\ }\textbf {\bibinfo {volume} {13}},\ \bibinfo {pages} {1586773} (\bibinfo {year} {2025})}\BibitemShut {NoStop}%
\bibitem [{\citenamefont {Henneberger}(1968)}]{Henneberger}%
  \BibitemOpen
  \bibfield  {author} {\bibinfo {author} {\bibfnamefont {W.~C.}\ \bibnamefont {Henneberger}},\ }\bibfield  {title} {\bibinfo {title} {Perturbation method for atoms in intense light beams},\ }\href {https://doi.org/10.1103/PhysRevLett.21.838} {\bibfield  {journal} {\bibinfo  {journal} {Phys. Rev. Lett.}\ }\textbf {\bibinfo {volume} {21}},\ \bibinfo {pages} {838} (\bibinfo {year} {1968})}\BibitemShut {NoStop}%
\bibitem [{\citenamefont {Su}\ \emph {et~al.}(1979)\citenamefont {Su}, \citenamefont {Schrieffer},\ and\ \citenamefont {Heeger}}]{Su1979}%
  \BibitemOpen
  \bibfield  {author} {\bibinfo {author} {\bibfnamefont {W.~P.}\ \bibnamefont {Su}}, \bibinfo {author} {\bibfnamefont {J.~R.}\ \bibnamefont {Schrieffer}},\ and\ \bibinfo {author} {\bibfnamefont {A.~J.}\ \bibnamefont {Heeger}},\ }\bibfield  {title} {\bibinfo {title} {Solitons in polyacetylene},\ }\href {https://doi.org/10.1103/PhysRevLett.42.1698} {\bibfield  {journal} {\bibinfo  {journal} {Phys. Rev. Lett.}\ }\textbf {\bibinfo {volume} {42}},\ \bibinfo {pages} {1698} (\bibinfo {year} {1979})}\BibitemShut {NoStop}%
\bibitem [{\citenamefont {Li}\ \emph {et~al.}(2017)\citenamefont {Li}, \citenamefont {Lin}, \citenamefont {Zhang},\ and\ \citenamefont {Song}}]{Li2017}%
  \BibitemOpen
  \bibfield  {author} {\bibinfo {author} {\bibfnamefont {C.}~\bibnamefont {Li}}, \bibinfo {author} {\bibfnamefont {S.}~\bibnamefont {Lin}}, \bibinfo {author} {\bibfnamefont {G.}~\bibnamefont {Zhang}},\ and\ \bibinfo {author} {\bibfnamefont {Z.}~\bibnamefont {Song}},\ }\bibfield  {title} {\bibinfo {title} {Topological nodal points in two coupled {Su-Schrieffer-Heeger} chains},\ }\href {https://doi.org/10.1103/PhysRevB.96.125418} {\bibfield  {journal} {\bibinfo  {journal} {Phys. Rev. B}\ }\textbf {\bibinfo {volume} {96}},\ \bibinfo {pages} {125418} (\bibinfo {year} {2017})}\BibitemShut {NoStop}%
\bibitem [{\citenamefont {Dal~Lago}\ \emph {et~al.}(2015)\citenamefont {Dal~Lago}, \citenamefont {Atala},\ and\ \citenamefont {Torres}}]{dalLago2015floquet}%
  \BibitemOpen
  \bibfield  {author} {\bibinfo {author} {\bibfnamefont {V.}~\bibnamefont {Dal~Lago}}, \bibinfo {author} {\bibfnamefont {M.}~\bibnamefont {Atala}},\ and\ \bibinfo {author} {\bibfnamefont {L.~F.}\ \bibnamefont {Torres}},\ }\bibfield  {title} {\bibinfo {title} {Floquet topological transitions in a driven one-dimensional topological insulator},\ }\href {https://doi.org/10.1103/PhysRevA.92.023624} {\bibfield  {journal} {\bibinfo  {journal} {Phys. Rev. A}\ }\textbf {\bibinfo {volume} {92}},\ \bibinfo {pages} {023624} (\bibinfo {year} {2015})}\BibitemShut {NoStop}%
\bibitem [{\citenamefont {Borja}\ \emph {et~al.}(2022)\citenamefont {Borja}, \citenamefont {Gutiérrez},\ and\ \citenamefont {López}}]{Borja2022}%
  \BibitemOpen
  \bibfield  {author} {\bibinfo {author} {\bibfnamefont {C.}~\bibnamefont {Borja}}, \bibinfo {author} {\bibfnamefont {E.~D.}\ \bibnamefont {Gutiérrez}},\ and\ \bibinfo {author} {\bibfnamefont {A.}~\bibnamefont {López}},\ }\bibfield  {title} {\bibinfo {title} {Emergence of floquet edge states in the coupled su-schrieffer-heeger model},\ }\href {https://doi.org/10.1088/1361-648X/ac5865} {\bibfield  {journal} {\bibinfo  {journal} {J. Phys. Condens. Matter}\ }\textbf {\bibinfo {volume} {34}},\ \bibinfo {pages} {205701} (\bibinfo {year} {2022})}\BibitemShut {NoStop}%
\bibitem [{\citenamefont {Shirley}(1965)}]{Shirley1965}%
  \BibitemOpen
  \bibfield  {author} {\bibinfo {author} {\bibfnamefont {J.~H.}\ \bibnamefont {Shirley}},\ }\bibfield  {title} {\bibinfo {title} {Solution of the schr\"odinger equation with a hamiltonian periodic in time},\ }\href {https://doi.org/10.1103/PhysRev.138.B979} {\bibfield  {journal} {\bibinfo  {journal} {Phys. Rev.}\ }\textbf {\bibinfo {volume} {138}},\ \bibinfo {pages} {B979} (\bibinfo {year} {1965})}\BibitemShut {NoStop}%
\bibitem [{\citenamefont {Grifoni}\ and\ \citenamefont {H\"anggi}(1998)}]{Grifoni1998}%
  \BibitemOpen
  \bibfield  {author} {\bibinfo {author} {\bibfnamefont {M.}~\bibnamefont {Grifoni}}\ and\ \bibinfo {author} {\bibfnamefont {P.}~\bibnamefont {H\"anggi}},\ }\bibfield  {title} {\bibinfo {title} {Driven quantum tunneling},\ }\href {https://doi.org/10.1016/S0370-1573(98)00022-2} {\bibfield  {journal} {\bibinfo  {journal} {Phys. Rep.}\ }\textbf {\bibinfo {volume} {304}},\ \bibinfo {pages} {229} (\bibinfo {year} {1998})}\BibitemShut {NoStop}%
\bibitem [{\citenamefont {Giovannini}\ and\ \citenamefont {Hübener}(2019)}]{Giovannini_2020}%
  \BibitemOpen
  \bibfield  {author} {\bibinfo {author} {\bibfnamefont {U.~D.}\ \bibnamefont {Giovannini}}\ and\ \bibinfo {author} {\bibfnamefont {H.}~\bibnamefont {Hübener}},\ }\bibfield  {title} {\bibinfo {title} {Floquet analysis of excitations in materials},\ }\href {https://doi.org/10.1088/2515-7639/ab387b} {\bibfield  {journal} {\bibinfo  {journal} {J. Phys. Mater.}\ }\textbf {\bibinfo {volume} {3}},\ \bibinfo {pages} {012001} (\bibinfo {year} {2019})}\BibitemShut {NoStop}%
\end{thebibliography}%

\end{document}